\documentclass{article}

\usepackage{graphicx}
\usepackage{subfig}
\usepackage{natbib}
\usepackage{amsmath}
\usepackage{amsfonts}
\usepackage{algorithm}
\usepackage{algpseudocode}
\usepackage{hyperref}

\title{Pre-processing for approximate Bayesian computation in image analysis}
\author{Matthew T. Moores\thanks{Mathematical Sciences School, Queensland University of Technology, GPO Box 2434, Brisbane QLD 4001, Australia}~\thanks{Current address: Department of Statistics, University of Warwick, Coventry, UK}         \and Christopher C. Drovandi\footnotemark[1] \and
        Kerrie Mengersen\footnotemark[1] \and
Christian P. Robert\thanks{CEREMADE, Universit\'e Paris Dauphine, 75775 Paris cedex 16, France}~\thanks{CREST, INSEE, France}~\thanks{Department of Statistics, University of Warwick, Coventry, UK}
}

\begin{document}

\maketitle

\begin{abstract}
Most of the existing algorithms for approximate Bayesian computation (ABC) assume that it is feasible to simulate pseudo-data from the model at each iteration. However, the computational cost of these simulations can be prohibitive for high dimensional data.  An important example is the Potts model, which is commonly used in image analysis.  Images encountered in real world applications can have millions of pixels, therefore scalability is a major concern. We apply ABC with a synthetic likelihood to the hidden Potts model with additive Gaussian noise. Using a pre-processing step, we fit a binding function to model the relationship between the model parameters and the synthetic likelihood parameters. Our numerical experiments demonstrate that the precomputed binding function dramatically improves the scalability of ABC, reducing the average runtime required for model fitting from 71 hours to only 7 minutes. We also illustrate the method by estimating the smoothing parameter for remotely sensed satellite imagery. Without precomputation, Bayesian inference is impractical for datasets of that scale.
\end{abstract}

\section{Introduction}
For many important problems, the computational cost of approximate Bayesian computation (ABC) is dominated by simulation of pseudo-data. This is particularly the case for the Potts model, for which an ABC algorithm was developed by \citet{Grelaud2009}. For real world applications in image analysis, the dimension of the state vector can correspond to millions of pixels. The distribution of these states is highly correlated, requiring algorithms such as \citet{Swendsen1987} to simulate efficiently from the generative model. In his comparison of sequential Monte Carlo (SMC-ABC) with a particle Markov chain Monte Carlo (PMCMC) algorithm, \citet{Everitt2012} found in both cases that the computational requirements were dominated by simulation of this state vector.

Adaptive ABC algorithms have been developed to reduce the number of iterations required through more efficient exploration of the posterior. Various particle-based methods have been proposed by \citet{Sisson2007,Beaumont2009,Toni2009,Drovandi2011,Jasra2012,Filippi2013} and \citet{Sedki2013}. The SMC-ABC algorithm of \citet{DelMoral2012} uses multiple replicates of the summary statistics for each particle, which accounts for heteroskedasticity in the relationship between the statistics and the parameter values. This algorithm also has the advantage that computation of the importance weights is linear in the number of particles. Both the ABC tolerance $\epsilon$ and the Metropolis-Hastings (MH) proposal bandwidth $\sigma^2_\mathrm{MH}$ are selected adaptively, which reduces the need to tune SMC-ABC for specific applications.

Regression adjustment for ABC was introduced by \citet{Beaumont2002}, who post-processed the ABC output by fitting a local linear regression. By modelling the relationship between the simulated parameter values and the corresponding summary statistics, they improved the estimate of the posterior density even with large values of the ABC tolerance. \citet{Blum2010a} took a similar approach, except that they used a nonlinear, heteroskedastic regression. They then performed a second ABC run, using the estimate of the posterior to draw parameter values.

The idea proposed in this paper has a strong connection with indirect inference \citep[e.g.][]{Gourieroux1993}.  We assume that the underlying (intractable) distribution of the summary statistic of interest can be approximated well by an alternative parametric model with a tractable distribution.  This so-called auxiliary model contains a different set of parameters.  The indirect inference nature of our method requires learning the relationship between the parameters of the auxiliary model and the true model, often referred to as the mapping or binding function.  \citet{Wood2010} assume a normal parametric model with mean and covariance parameters.  Our application involves a single summary statistic and we adopt the approach of \citet{Wood2010} and use a normal distribution with a mean and a variance parameter.  \citet{Wood2010} learn the auxiliary parameters for each value of the true parameter proposed during a MCMC run by simulating independently a large collection (say of size $M$) of summary statistics from the true parameter and estimating the corresponding auxiliary parameters based on the sample moments of the generated summary statistics.  We note that the resulting target distribution of the method depends on the choice of $M$ \citep{Drovandi2014}.  Our approach differs in that we estimate the binding function prior to running our Bayesian algorithm via producing model simulations over a pre-defined grid (referred to here as the precomputation step).  We attempt to recover the true mapping between the mean and variance auxiliary parameters with the true parameters (i.e.\ the limit as $M \rightarrow \infty$) by using non-parametric regression techniques. An additional advantage is that the output of the precomputation can be reused for multiple datasets that share the same parameter space.

Despite superficial similarities with the ``accurate ABC" method of \citet{Ratmann2014}, where an auxiliary model is
also constructed over a sufficient statistic, our approach cannot be seen as a special case of theirs as accurate ABC
requires repeated observations of the summary statistics and consistency of the auxiliary model for its calibration
step. We do not claim such proximity with the true posterior distribution and doubt it can be achieved for the models we
consider below.  Even though we are in the favourable case when the summary statistic is sufficient, most assumptions in \citet{Ratmann2014} do not apply to our setting. The construction of the binding function also highly differs in both
motivations and complexity. We nonetheless agree that those different approaches of \citet{Wood2010,Drovandi2011a,Cabras2014,Ratmann2014}, as well as ours, all relate in spirit to the original
indirect inference method of \citet{Gourieroux1993}. The reader is referred to \citet{Drovandi2014} for more details about the parametric auxiliary model approach adopted in this paper.

The rest of the paper is organised as follows. The ABC rejection sampler and SMC-ABC are reviewed in Sect.~\ref{s:abc}. Precomputation of the binding function is described in Sect.~\ref{s:method}.  In Sect.~\ref{s:potts}, we illustrate how this method can be applied to the hidden Potts model with additive Gaussian noise. Sect.~\ref{s:results} contains results from a simulation study as well as real imaging data from the Landsat 7 satellite. The article concludes with a discussion.

\section{Approximate Bayesian Computation}
\label{s:abc}
The ABC rejection sampler introduced by \citet{Pritchard1999} draws values of the parameter from its prior distribution $\pi(\theta)$, then simulates pseudo-data $\mathbf{x}$ from the model. One or more summary statistics $s(\mathbf{x})$ are calculated from the pseudo-data and compared to the values of those statistics in the observed data, $s(\mathbf{y})$. If the difference between the statistics is within the ABC tolerance threshold, then the proposed parameter is accepted. 

The simulation of pseudo-data from the generative model is typically the most computationally intensive step in this process. The ABC rejection sampler works best when prior information about the distribution of the parameter is available. The acceptance rate under a sparse or uninformative prior can be extremely low, requiring many pseudo-datasets to be generated for each parameter value that is accepted. The ABC tolerance is a tunable parameter, since a large tolerance means a higher acceptance rate but also increases the error in the estimate of the posterior distribution. If the summary statistic is sufficient for the parameter, then the distribution of the samples approaches the true posterior as $\epsilon$ approaches zero. However, the number of samples that are rejected also increases, to the point that almost none are accepted. Adaptive ABC methods have been developed to address this inefficiency.

\subsection{Sequential Monte Carlo}
\label{s:smc_abc}
The SMC-ABC algorithm of \citet{DelMoral2012} evolves a set of $N$ parameter values, known as particles, through a sequence of $T$ target distributions:
$$
\pi_t(\theta, \mathbf{x}_{1 \dots M} | \mathbf{y}) \propto \frac{\sum_{m=1}^M \mathbb{I}\left(\Delta(\mathbf{x}_m) < \epsilon_t\right)}{M} \left( \prod_{m=1}^M p(\mathbf{x}_m | \theta) \right) \pi(\theta)
$$
where $\epsilon_t$ is the ABC tolerance threshold such that $\epsilon_0 \ge \epsilon_1 \ge \dots \ge \epsilon_T$, $\mathbf{x}_{1 \dots M}$ is a set of $M$ replicates of the pseudo-data that are generated for each particle, $p(\mathbf{y}|\theta)$ is the likelihood, $\mathbb{I}(\cdot)$ is the indicator function, and $\Delta(\mathbf{x})$ is the distance between the summary statistics of the pseudo-data and the observed data, $\| s(\mathbf{x}) - s(\mathbf{y}) \|$, for an appropriate norm. The algorithm involves four major stages: initialisation, adaptation, resampling, and mutation.
%%
%\begin{algorithm}
%\begin{algorithmic}[1]
%\Statex {\em Initialisation:}
%\State $t \gets 0$, $\epsilon_0 \gets \infty$
%\State Draw $\theta_{i,0} \sim \pi(\theta) \; \forall i \in 1 \dots N$
%\State Generate $\vec{x}_{i, m, 0} \sim f(\cdot|\theta_{i,0}) \; \forall i \in 1 \dots N,  \; \forall m \in 1 \dots M$
%\State $w_{i,0} \gets \frac{1}{N} \; \forall i \in 1 \dots N$
%\Repeat
%\State $t \gets t+1$
%\Statex {\em Adaptation:}
%\State Update $\epsilon_t$ by solving (\ref{eq:tolerance})
%\State Update $w_{i,t} \; \forall i \in 1 \dots N$ according to  (\ref{eq:weight})
%\Statex {\em Resampling:}
%\If{$ESS_t < N_\mathrm{min}$}
%\State Resample $\theta_{i,t} \; \forall i \in 1 \dots N$
%\State $w_{i,t} \gets \frac{1}{N} \; \forall i \in 1 \dots N$
%\EndIf
%\Statex {\em Mutation:}
%\State $n_\mathrm{accept} \gets 0$
%\ForAll{$i \in 1 \dots N$}
%\If{$w_{i,t} > 0$}
%\State Draw $\theta_i' \sim q_t(\cdot|\theta_{\cdot,t-1})$
%\State Generate $\vec{x}_{i,m}' \sim f(\cdot|\theta_i')  \; \forall m \in 1 \dots M$
%\State Calculate $\rho_i$ according to  (\ref{eq:mh_rho})
%\State Draw $u \sim \mathcal{U}(0,1)$
%\If{$u < \rho_i$}
%\State $\left(\theta_{i,t}, \vec{x}_{i, \cdot, t}\right) \gets \left(\theta_i', \vec{x}_{i,\cdot}'\right)$
%\State $n_\mathrm{accept} \gets n_\mathrm{accept} + 1$
%\Else
%\State $\left(\theta_{i,t}, \vec{x}_{i, \cdot, t}\right) \gets \left(\theta_{i,t-1}, \vec{x}_{i, \cdot, t-1}\right)$
%\EndIf
%\EndIf
%\EndFor
%\Until{$\frac{n_\mathrm{accept}}{N} < 0.015$ {\bf or} $\epsilon_t = 0$}
%\end{algorithmic}
%\caption{SMC-ABC}
%\label{alg:smc-abc}
%\end{algorithm}

\paragraph{Initialisation} The algorithm is initialised at $t=0$ by drawing a population of particles $\theta_{i,t}$ from the prior, where $i \in 1 \dots N$. Each particle is associated with $M$ replicates of the summary statistics calculated from pseudo-data. The generation of multiple sets of pseudo-data $\mathbf{x}_{i,m,t}$ for each particle increases the computational cost relative to other ABC methods, but it better handles the situation where there is sizeable variability in the value of the summary statistic for a given parameter. This is the case for the hidden Potts model, as we explain in Sect.~\ref{s:potts}.

\paragraph{Adaptation} At each iteration, the particles are assigned importance weights based on the following formula:
\begin{equation}
\label{eq:weight}
w_{i,t} \propto w_{i,t-1} \frac{\sum_{m=1}^{M} \mathbb{I}\left( \Delta(\mathbf{x}_{i, m, t-1}) < \epsilon_t\right)}{\sum_{m=1}^{M} \mathbb{I}\left( \Delta(\mathbf{x}_{i, m, t-1}) < \epsilon_{t-1}\right)}
\end{equation}
These weights are normalised so that $\sum_{i=1}^N w_{i,t} = 1$. These weights gradually degenerate over successive iterations, which is measured by the effective sample size \citep[$ESS$;][pp. 34--36]{Liu2001}:
\begin{equation}
\label{eq:ess}
ESS_t = \left(\sum_{i=1}^N w_{i,t}^2\right)^{-1}
\end{equation}
The ABC tolerance $\epsilon_t$ is updated adaptively according to the desired rate $\alpha \in (0,1)$ of the reduction in the $ESS$:
\begin{equation}
\label{eq:tolerance}
ESS_t = \alpha ESS_{t-1}
\end{equation}
This equation must be solved iteratively, e.g. by interval bisection, since $ESS_t$ depends on the weights $w_{\cdot,t}$, which in turn depend on $\epsilon_t$ according to (\ref{eq:weight}).

\paragraph{Resampling} If $ESS_t$ falls below a threshold value $N_\mathrm{min}$ then the particles are all resampled. The new population of $N$ particles can either be drawn from a multinomial distribution with weights $\lambda_i = w_{i,t}$ or more complicated schemes can be used. \citeauthor{DelMoral2012} employed the systematic resampling scheme of \citet{Kitagawa1996}. Once the particles have been resampled, all of the importance weights are set to $N^{-1}$ and thus the $ESS$ is equal to $N$.

\paragraph{Mutation} Finally, the particles with nonzero weight are updated using a random walk proposal $q_t(\theta'|\theta_{t-1})$. The bandwidth $\sigma^2_\mathrm{MH}$ can be chosen adaptively using an importance sampling approximation of the variance of $\theta$ under $\pi_{t-1}(\theta|\mathbf{y})$, as in \citet{Beaumont2009}. The pseudo-data is also updated using $q(\mathbf{x}'|\theta')$ and jointly accepted with probability $\min(1,\rho_i)$ according to the MH acceptance ratio:
\begin{equation}
\label{eq:mh_rho}
\rho_i = \frac{\sum_{m=1}^{M} \mathbb{I}\left( \Delta(\mathbf{x}_{i,m}') < \epsilon_t\right)}{\sum_{m=1}^{M} \mathbb{I}\left( \Delta(\mathbf{x}_{i, m, t-1}) < \epsilon_{t}\right)} \frac{q_t(\theta_{i,t-1}|\theta_i')}{q_t(\theta_i'|\theta_{i,t-1})} \frac{\pi(\theta_i')}{\pi(\theta_{i,t-1})}
\end{equation}

\section{Precomputation of the Summary Statistic}
\label{s:method}
Model fitting with ABC can be decomposed into two separate stages: learning about the summary statistic, given the parameter $\left(s(\mathbf{x})|\theta\right)$; and choosing parameter values, given the summary statistic $\left(\theta|s(\mathbf{y})\right)$. The first stage is achieved by simulating pseudo-data from the generative model in a so-called precomputation step (discussed below), while the second is achieved via a Bayesian algorithm that utilises the output of the precomputation. In the case of latent models there are additional complications that will be discussed further in Sect.~\ref{s:noise}.  To ease the exposition of this section, we assume the data $\mathbf{y}$ are observed without error.

The precomputation step involves simulating pseudo-data for fixed values of the parameter.  Firstly we approximate the intractable distribution of the summary statistic, $f(s(\mathbf{y})|\theta)$, with an alternative parametric model that has a tractable likelihood function, $f_A(s(\mathbf{y})|\phi(\theta))$, which contains a different set of parameters $\phi$.  There is a strong connection with indirect inference (e.g.\ \citet{Gourieroux1993}) as our method requires learning the mapping between $\theta$ and $\phi$, $\phi(\theta)$, often referred to as the binding function.  

\citet{Wood2010} adopts this approach and uses the auxiliary likelihood within a MCMC algorithm.  For each proposed value of $\theta$, a set of $M$ independent pseudo-datasets are generated from the true model, $\mathbf{x}_{1:M} = (\mathbf{x}_1,\ldots,\mathbf{x}_M)$, and the corresponding summary statistics are constructed, $s_{1:M} = (s(\mathbf{x}_1),\ldots,s(\mathbf{x}_M))$, which can be viewed as an independent and identically distributed sample from $s(\mathbf{x})|\theta$.  The auxiliary model is then fit (using maximum likelihood or the method moments for example) to this sample in order to estimate the parameter $\phi$.  We denote this estimate of the mapping $\phi(\theta)$ as $\phi(\theta,s_{1:M})$. \citet{Drovandi2014} note that the target distribution of this method depends on the value of $M$, and if the auxiliary model chosen is suitable, it is desirable to take $M$ as large as possible.

The approach that we use to learn the mapping is different to that in \citet{Wood2010}.  Here we define a (not necessarily regular) grid over the parameter space $\Theta$.  For each $\theta$ within the grid a set of summary statistics, $s_{1:M}$, are generated from the true model and the corresponding auxiliary parameter is estimated, $\phi(\theta,s_{1:M})$.  The next step is to use non-parametric regression techniques in order to smooth out the effect due to a finite choice of $M$ and to obtain a direct approximation of the mapping, which we denote $\hat{\phi}(\theta)$.  Here each component of $\phi$ is regressed on the model parameter $\theta$.   These non-parametric regression models can be used to predict the mapping for $\theta$ values not present in the grid.  We refer to this procedure for estimating the mapping as the precomputation step.  Our approach, in addition to mitigating the effect of $M$, has the advantage of being very useful when fitting the same model to multiple datasets because the output of the precomputation can be reused, thus amortising its computational cost. For alternative approaches to combining nonparametric regression models with ABC, see the regression adjustment method of \citet{Blum2010a} and the meta-modelling or emulation approach of \citep{Wilkinson2014}.

Our application involving the Potts model only has a single summary statistic and we follow \citet{Wood2010} and assume a normal distribution for $f_A$ with mean $\mu$ and variance $\sigma^2$.  Therefore our precomputation step requires the generation of $M$ pseudo-datasets for each value of $\beta$ across a grid, recording the sample mean and standard deviation of the summary statistic (i.e.\ estimates of the auxiliary parameters $\mu$ and $\sigma$) and then applying two separate non-parametric regressions to estimate the mappings $\mu(\beta)$ and $\sigma(\beta)$.  It is important to note that for reasonable size images it is not computationally feasible to generate perfect samples from the Potts model.  Instead we use an MCMC algorithm to generate a set of $M$ correlated pseudo-datasets and use these to calculate $s_{1:M}$ (see Sect.~\ref{s:pre_Potts} for specific details of this for the Potts model).  

There are several approximations induced by our method.  The first is associated with replacing $f$ with $f_A$.  Despite the summary statistic of the Potts model being discrete, our choice of the normal distribution is appropriate due to the size of the images being analysed.  The second source of approximation arises from the estimated mapping, $\hat{\phi}(\theta)$, due to the non-parametric regressions.  However, the quality of the constructed mapping can be assessed visually and/or by standard data analytic techniques.  For convenience in this paper we also introduce a third and seemingly unnecessary approximation effect.  The natural implementation, used in \citet{Wood2010} and \citet{Drovandi2014} for example, uses the auxiliary likelihood estimate $f_A(s(\mathbf{y})|\hat{\phi}(\theta))$ directly in a Bayesian algorithm and avoids any comparison of observed and simulated summary statistics (that is, specification of an ABC tolerance $\epsilon$ is not required).  In this paper we use a more traditional ABC approach and draw pseudo summary statistics from $f_A(s(\cdot)|\hat{\phi}(\theta))$ for comparison with the observed data (and thus require $\epsilon$).  We adopt this approach here for two reasons.  The first is for ease of implementation; having already developed an SMC-ABC algorithm, all that is required is to replace the (expensive) simulation of pseudo-data from the model with the (cheap) summary statistic draw from the auxiliary model.  The second motivation for our implementation approach is that a more direct comparison of the computational cost can be made between the SMC-ABC method with and without the precomputation step.

\section{Hidden Potts Model}
\label{s:potts}
We illustrate our method using a hidden Potts model. The Potts model is a Markov random field with discrete states $z \in 1\dots k$. It is defined in terms of its conditional probabilities:
  \begin{equation}
  \label{eq:Potts}
p(z_i|z_{i\sim \ell}, \beta) = \frac{\exp\left\{\beta\sum_{i \sim \ell}\delta(z_i,z_\ell)\right\}}{\sum_{j=1}^k \exp\left\{\beta\sum_{i \sim \ell}\delta(j,z_\ell)\right\}}
  \end{equation}
where $i \in 1 \dots n$ are the nodes in the image lattice, also known as pixels, $\beta$ is a scale parameter known as the inverse temperature, $i \sim \ell$ are the neighbouring pixels of $i$, and $\delta(u,v)$ is the Kronecker delta function. In this paper we use the first-order neighbourhood, so $i \sim \ell$ are the four pixels immediately adjacent to an internal node of the lattice. Pixels on the image boundary have less than four neighbours.

The inverse temperature parameter governs the strength of spatial association. A value of zero corresponds with spatial independence, while values greater than zero increase the probability of adjacent neighbours having the same state.  The full conditional distribution of the inverse temperature is given by
  \begin{equation}
  \label{eq:beta}
  p(\beta|\vec{z}) = \mathcal{C}(\beta)^{-1} \pi(\beta) \exp\left\{ \beta \mathrm{S}(\mathbf{z}) \right\}
  \end{equation}
  where $\mathcal{C}(\beta)$ is an intractable normalising constant. It involves a sum over all $k^n$ possible combinations of the labels $\mathbf{z} \in \mathcal{Z}$:
  \begin{equation}
  \label{eq:norm}
\mathcal{C}(\beta) = \sum_{\mathbf{z} \in \mathcal{Z}} \exp\left\{\beta \mathrm{S}(\mathbf{z})\right\}
  \end{equation}

\begin{figure*}
  \subfloat[Expectation of $\mathrm{S}(\vec{z})$]{\label{f:exact_exp_n} \includegraphics[scale=0.33]{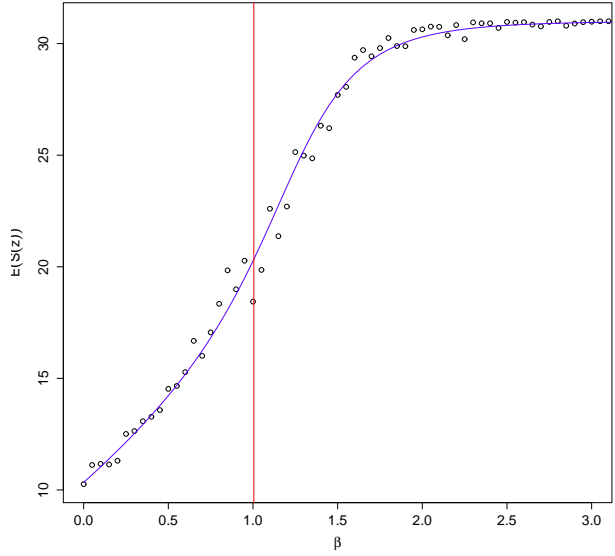}}
\qquad
  \subfloat[Standard deviation of $\mathrm{S}(\vec{z})$]{\label{f:exact_var_n} \includegraphics[scale=0.33]{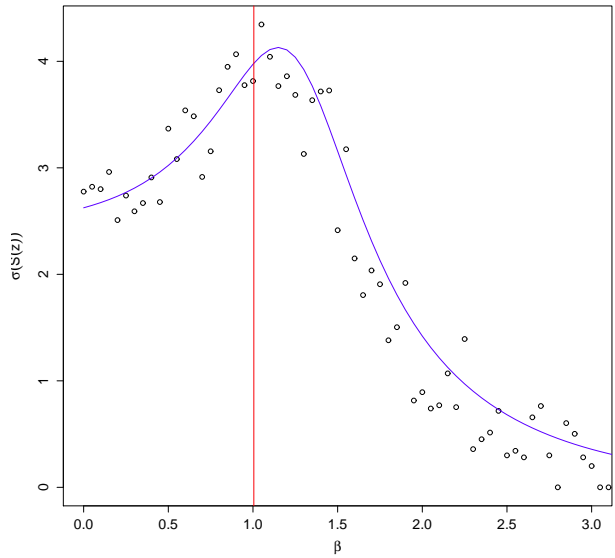}}
\caption{Comparison of the exact distribution of the Potts model for a trivial image ($n=20$, $k=3$) with approximations using the Swendsen-Wang algorithm. The continuous, curved line is the exact value and the points are the approximations. The vertical line marks the critical value of $\beta$.}
\label{f:exact_beta}
\end{figure*}

A sufficient statistic is available for this model since it belongs to the exponential family, as noted by \citet{Grelaud2009}. If $\mathcal{E}$ is the set of all unique neighbour pairs, or edges in the image lattice, then the sufficient statistic is
  \begin{equation}
  \label{eq:stat}
\mathrm{S}(\mathbf{z}) = \sum_{i \sim \ell \in \mathcal{E}} \delta(z_i,z_\ell)
  \end{equation}
 Thus, this statistic represents the total number of like neighbor pairs in the image. As $\beta$ approaches infinity, all of the pixels in the image are almost surely assigned the same label, thus the expectation of $\mathrm{S}(\mathbf{z})$ approaches the total number of edges $|\mathcal{E}|$ asymptotically, while the variance approaches zero. When $\beta=0$, the probability of any pair of neighbours being assigned the same label follows an independent Bernoulli distribution with $p = k^{-1}$, thus $\mathrm{S}(\mathbf{z})$ follows a Binomial distribution with expectation $|\mathcal{E}| /k$ and variance $|\mathcal{E} \; | k^{-1} (1 - k^{-1})$. The distribution of $\mathrm{S}(\mathbf{z})$ changes smoothly between these two extremes, as illustrated by Fig.~\ref{f:exact_beta}, but its computation is intractable for nontrivial images. The expectations and standard deviations in Fig.~\ref{f:exact_beta} were calculated for $k=3$  unique labels and $n=20$ pixels, much less than required for any practical application.

The maximum variance of $\mathrm{S}(\mathbf{z}) | \beta$, which corresponds to the steepest gradient in the expectation, occurs near the critical temperature. This is the point at which the Potts model transitions from a disordered to an ordered state. The phase transition behaviour has analogies in physical systems, such as the Curie temperature in ferromagnetic materials. When $\beta > \beta_\mathrm{crit}$, the values of the labels begin to exhibit long-range dependence and coalesce into clusters of similar values. \citet{Potts1952} showed that this critical point can be calculated exactly for a two-dimensional regular lattice by $\beta_\mathrm{crit} = \log\{1 + \sqrt{k}\}$, so $\beta_\mathrm{crit} \approx 1.005$ for $k=3$ and $\beta_\mathrm{crit} \approx 1.238$ for $k=6$. The nonlinearity and heteroskedasticity evident in Fig.~\ref{f:exact_beta} will need to be accounted for in our choice of binding function $\hat\phi(\beta)$.

\subsection{Precomputation of $S(z)$}
\label{s:pre_Potts}
\begin{figure*}
  \subfloat[Estimate of the expectation $\mathrm{E}\{\mathrm{S}(\mathbf{z}) | \beta\}$]{\label{f:simSz_n125k3_mu} \includegraphics[scale=0.34]{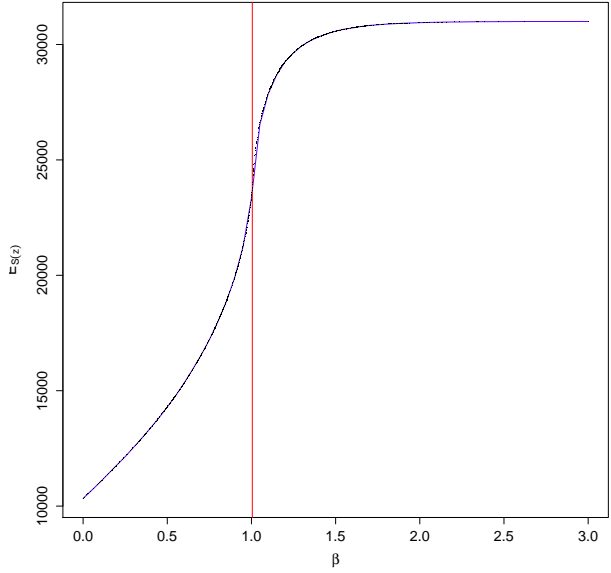}}
\qquad
  \subfloat[Estimate of the standard deviation $\sigma\{\mathrm{S}(\mathbf{z}) | \beta\}$ ]{\label{f:simSz_n125k3_sd} \includegraphics[scale=0.34]{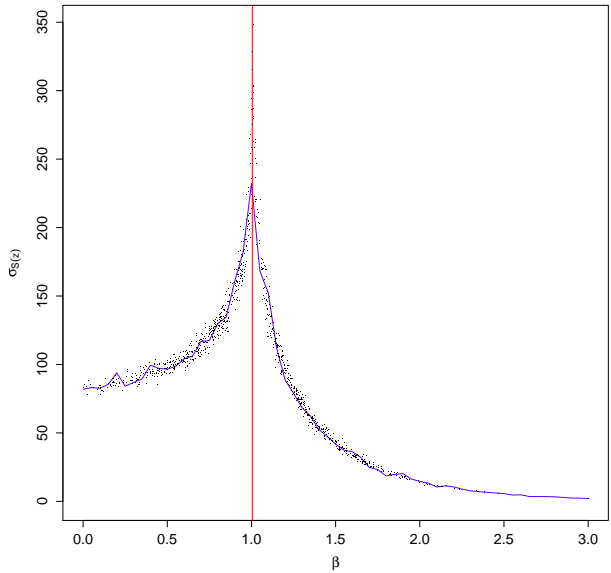}}
\caption{Approximation of the expectation and standard deviation of the sufficient statistic for the Potts model with $n=125\times 125$ and $k=3$. The continuous line shows linear interpolation for 61 values of $\mathrm{S}(\vec{z}) | \beta$, while the points are 987 values approximated using Swendsen-Wang.}
\label{f:simSz}
\end{figure*}
Since it is impossible to sample from $\beta|\mathbf{z}$ directly, we use ABC methods. This requires simulating pseudo-data from the Gibbs distribution $\pi(z_i|z_{i\sim \ell}, \beta)$ defined by (\ref{eq:Potts}). It is difficult to simulate from this distribution because neighbouring pixels are highly correlated with each other, particularly for $\beta > \beta_\mathrm{crit}$, thus pixels can remain in the same state for many iterations. We use the algorithm of \citet{Swendsen1987}, which updates clusters of pixels simultaneously. The effect of the approximation error can be seen in Fig.~\ref{f:exact_beta} for a trivial image, where computation of the exact likelihood is feasible. Even though Swendsen-Wang is much less computationally intensive than perfect sampling, simulating pseudo-data remains expensive, which is why we approximate $f_A(\mathrm{S}(\mathbf{z}) | \hat\phi(\beta))$ offline using a pre-processing step.

To estimate the binding function for the Potts model, we use 1000 values of the inverse temperature, drawn from a truncated normal distribution $\beta \sim \mathcal{N}\left(\beta_\mathrm{crit}, (\beta_\mathrm{crit} / 2)^2\right) \mathbb{I}(\beta > 0)$. This concentrates computational effort in the vicinity of the critical region, as shown in Fig.~\ref{f:simSz}. The expectation and standard deviation have been approximated for a regular lattice with $n=125\times 125$ pixels and $k=3$, which corresponds to the simulation study in section~\ref{s:simulation}. For comparison, we show linear interpolation between 61 values of $\beta$ on a regularly spaced grid, so that the approximation error due to a finite sample size can be seen. The simpler binding function provides a very good fit for the distribution of the expected values, but there is a larger approximation error in the estimate of the standard deviation. This is particularly evident at the critical point, where the variance of the sufficient statistic is much larger than estimated. We found that a binding function based on only 61 values did not provide sufficient accuracy for our purposes.

We perform 1000 iterations of Swendsen-Wang for each value of $\beta$, discarding the first 500 as burn-in. The remaining $M=500$ iterations are used to compute the expectation and variance of the sufficient statistic. It should be noted that this operation is embarrassingly parallel, since the computation for each value of $\beta$ is completely independent. The results of this pre-processing step are stored in a matrix, which can then be used to fit the same model to multiple datasets. During model fitting, these pre-computed values are interpolated to obtain $\hat\mu(\beta')$ and $\hat\sigma(\beta')$, then the conditional distribution of the sufficient statistic is approximated by a Gaussian with these parameters.

\subsection{Additive Gaussian noise}
\label{s:noise}
\begin{figure}
\begin{center}
\includegraphics[width=\linewidth]{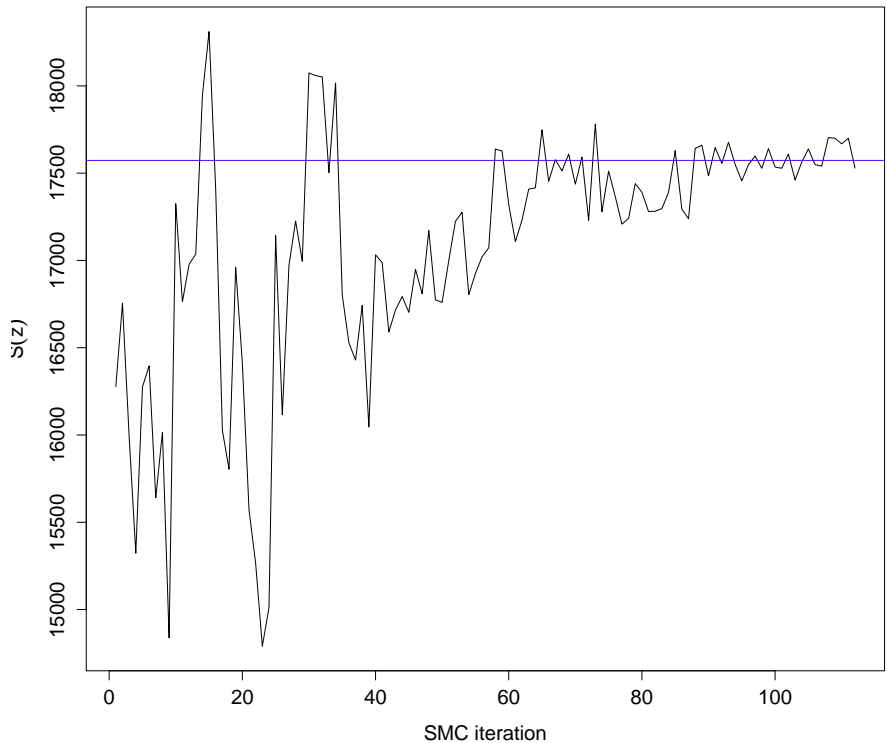}
\end{center}
\caption{Change in the value of the sufficient statistic according to the current distribution of $\pi_t(\beta|\mathbf{z})$. The horizontal line shows the true value of $\mathrm{S}(\mathbf{z})$.}
\label{f:smc-abc_S_z}
\end{figure}

In the hidden Potts model, the observed data $\mathbf{y}$ are independently distributed conditional on the latent labels $\mathbf{z}$. Under the assumption of additive Gaussian noise, the observation process is characterised by
\begin{equation}
\label{eq:obs}
y_i | z_i\!=\!j \quad\stackrel{iid}{\sim}\quad \mathcal{N}\left(\mu_j,\sigma_j^2\right)
\end{equation}
Since each unique label value corresponds to the mean and variance of a Normal distribution, this model can be viewed as a spatially-correlated generalisation of the mixture of Gaussians. 

When the parameters of these mixture components are unknown, they must be estimated as part of the model fitting procedure. The joint distribution of all random quantities is given by:
$$
p(\mathbf{y},\mathbf{z},\boldsymbol{\mu},\boldsymbol{\sigma^2},\beta) = p(\mathbf{y}|\mathbf{z},\mu,\sigma^2) \pi(\mu) \pi(\sigma^2) p(\mathbf{z}|\beta) \pi(\beta)
$$
where the vectors of noise parameters are $\boldsymbol{\mu} = (\mu_1, \dots, \mu_k)$ and $\boldsymbol{\sigma^2} = (\sigma_1^2, \dots, \sigma_k^2)$.

The conditional distribution of the latent labels is dependent on both the current likelihood of the data as well as the distribution of the particles:
\begin{equation}
\label{eq:post_z}
p(z_i | y_i, \boldsymbol\mu, \boldsymbol{\sigma^2}, \beta) = \frac{p(y_i; \mu_{z_i}, \sigma^2_{z_i})}{\sum_{j=1}^k p(y_i; \mu_j, \sigma^2_j)} \; p(z_i|z_{i\sim \ell}, \beta)
\end{equation}
where $p(y; \mu, \sigma^2)$ is the Gaussian likelihood and $p(z_i | z_{i\sim \ell}, \beta)$ is the Markov random field defined by (\ref{eq:Potts}).

This data augmentation approach can be problematic for SMC-ABC because it means that the summary statistic is a moving target, as shown in Fig.~\ref{f:smc-abc_S_z}. The summary statistic in (\ref{eq:stat}) is computed from the latent labels, therefore it only depends indirectly on the observed data through the noise parameters $\boldsymbol\mu, \boldsymbol{\sigma^2}$. Accurate posterior inference for $\beta$ is only possible to the extent that the pixel labels and the noise parameters have also been estimated correctly. It can take many SMC iterations for all of the components of the model to converge. This is the third stage of ABC estimation for latent models that was mentioned in section~\ref{s:method}.

One approach to overcoming this circular dependency would be to include the noise parameters $\mu_j,\sigma_j^2$ in the state vector for each SMC particle. Updating these parameters would require generating latent labels from (\ref{eq:post_z}), which would severely limit the scalability of the algorithm. Simulating from the distribution of $\pi(\mathbf{z} | \mathbf{y},\boldsymbol\mu, \boldsymbol{\sigma^2},\beta)$ is even more difficult than drawing pseudo-data from $\pi(\mathbf{z}|\beta)$ \citep{Hurn1997,Higdon1998}. It is simply infeasible to do this for each particle individually. Apart from the multiplication of computational cost by serveral orders of magnitude, there is also the issue of memory utilisation when updating particles in parallel. A single copy of the state vector can require several megabytes of memory, depending on the size of the image. Maintaining a copy of $\mathbf{z}$ for each thread would be impractical for massively parallel implementation of the algorithm.

We propose a pragmatic alternative that preserves the scalability of our algorithm, while enabling estimation of all of the components of the hidden Potts model. We run a single MCMC chain in conjunction with our SMC-ABC algorithm, although it would also be possible to maintain a small number of parallel chains. At the end of each SMC iteration, we draw a random sample from the current distribution of $\pi_t(\beta|\mathbf{z})$ according to the importance weights of the particles. These parameter values are used to update the pixel labels according to (\ref{eq:post_z}) by performing multiple iterations of chequerboard sampling~\citep[chap. 8]{Winkler2003}, one for each value of the parameter. After this sequence of updates, the new pixel labels reflect the distribution of $\beta$ at the current SMC iteration. It is important to note that it requires several iterations to make substantial changes to the distribution of $\mathbf{z}$, thus it is the aggregate effect of all of these updates that transitions from $\mathbf{z}|\pi_{t-1}(\beta)$ to $\mathbf{z}|\pi_t(\beta)$. The new state vector is used to calculate $\mathrm{S}(\mathbf{z})$, as well as the sufficient statistics of the noise parameters $\bar{y}_j | \mathbf{z}$ and $s_j^2 | \mathbf{z}$. We used 1000 MCMC iterations per iteration of SMC to produce the results in section~\ref{s:results}.

One downside of this method is that the MCMC chain can become stuck in a low-probability region of the state space. This is the cause of the outliers that are evident in Fig.~\ref{f:sim_results}. Once the Markov chain crosses the phase transition boundary of the Potts model from a disordered to an ordered state, the correlations between neighbouring pixels make the probability of transitioning back extremely low, irrespective of the values of $\beta$ that are used. To mitigate this problem, we initialize $\mathbf{z}, \boldsymbol\mu$ and $\boldsymbol{\sigma^2}$ at $\beta=0$ during the initialisation phase of our algorithm. This increases the probability that the starting value of $\mathrm{S}(\mathbf{z})$ will be below the critical region. 

\section{Illustration}
\label{s:results}
This section contains experimental results with synthetic data as well as real satellite imagery. In Sect.~\ref{s:simulation} we evaluate the performance and accuracy of our method for 20 images that have been simulated using the Swendsen-Wang algorithm for known values of $\beta$. We are thus able to compare the posterior distribution with the true parameter value. In Sect.~\ref{s:satellite} we demonstrate our method on an image obtained from the Landsat 7 satellite. This demonstrates that our pre-computation step enables inference with ABC for datasets of realistic size. We begin by calibrating our method against the SMC-ABC algorithm of \citet{DelMoral2012} as well as the approximate exchange algorithm \citep{Murray2006,Cucala2009,Friel2011}.

An \textsf{R} source package containing these algorithms is provided in Online Resource 1. Its computational engine is implemented in \textsf{C++} using RcppArmadillo \citep{Eddelbuettel2014} with OpenMP for parallelism. Reference implementations of SMC-ABC are available in the supplementary material accompanying \citet{DelMoral2012} and \citet{Everitt2012}. Source code for the approximate exchange algorithm has been provided by \citet{Friel2011} and \citet{Everitt2012}.

The elapsed times were recorded on 2.66GHz Intel Xeon processors. We used 8 parallel cores for fitting the model to each image and the precomputation was performed on a dual-CPU computer with 16 parallel cores. Memory usage varied depending on the number of pixels and the degree of parallelism. Approximate memory requirements for each computation are reported below.

\begin{figure}
\begin{center}
\includegraphics[width=\linewidth]{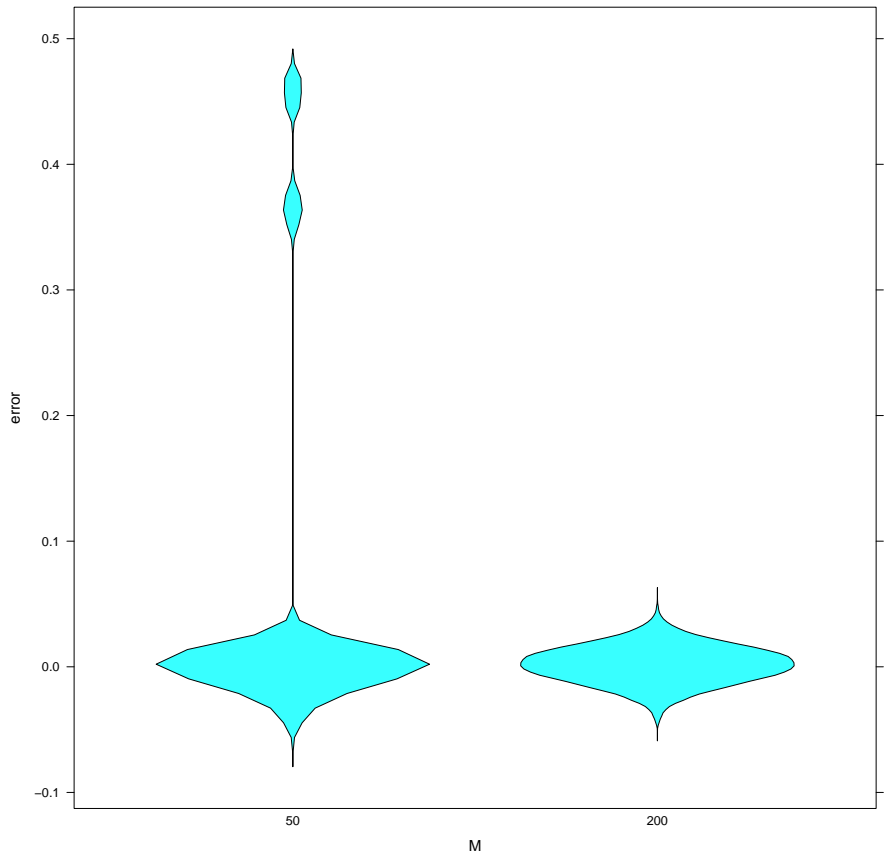}
\end{center}
\caption{Distribution of posterior sampling error for $\beta$, comparing SMC-ABC using $M=50$ replicates of the sufficient statistics with $M=200$.}
\label{f:beta_err}
\end{figure}
We fit the model to each image using $N=10,000$ SMC particles with $\alpha=0.97$, using residual resampling~\citep{Douc2005}. We drew $M=200$ summary statistics per particle from our precomputed binding function, but found that it was infeasible to simulate that much pseudo-data during model fitting since it took an average of 2 hours 45 minutes per SMC iteration. Instead, we used $M=50$ for SMC-ABC with pseudo-data, which resulted in runtimes that were more reasonable. This made a difference to the accuracy of the posterior, as shown in Fig.~\ref{f:beta_err}.

\begin{figure}
\begin{center}
\includegraphics[width=\linewidth]{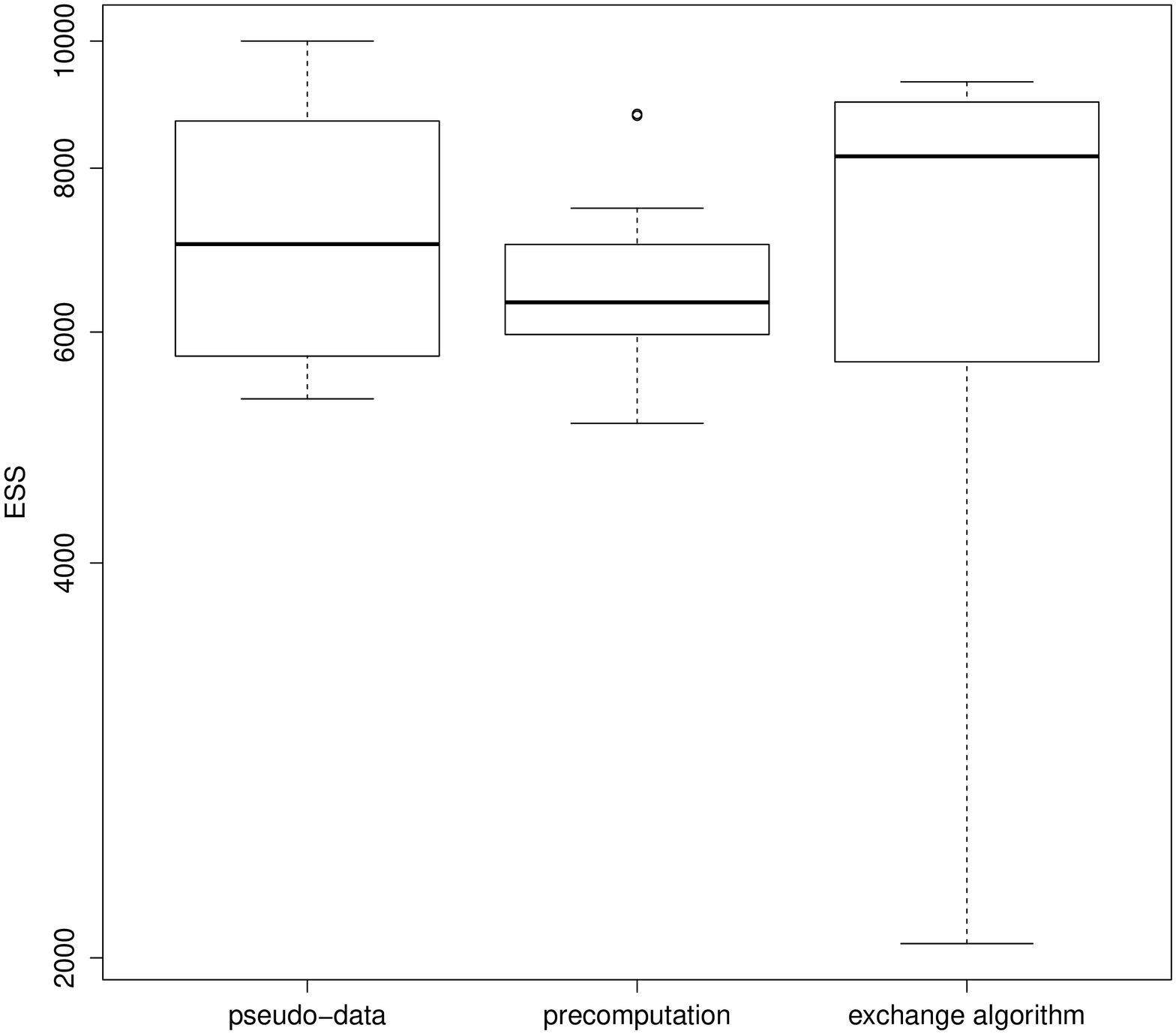}
\end{center}
\caption{Distribution of effective sample size (ESS) for model fitting with the exchange algorithm in comparison to SMC-ABC with pseudo-data or precomputed $\hat\phi(\beta)$.}
\label{f:ess}
\end{figure}
The exchange algorithm of \citet{Murray2006} is an exact method for intractable likelihoods, but only when it is feasible to use perfect sampling to simulate from $\pi(\mathbf{z} | \beta)$. Even then, many studies such as \citet{McGrory2009} have shown that this algorithm is very computationally intensive and takes longer as the value of $\beta$ increases. For this reason, \citet{Cucala2009} and \citet{Friel2011} replaced the perfect sampling step with 500 iterations of Gibbs sampling, hence creating an approximate exchange algorithm. The effect of this approximation on the accuracy of posterior inference has been studied by \citet{Everitt2012}. Since this is a MCMC method, the samples of $\pi(\beta|\mathbf{y})$ are correlated, reducing the effective sample size in comparison to the number of iterations. We found that 100,000 iterations were required to produce an ESS that was comparable with the SMC-ABC methods, as shown in Fig.~\ref{f:ess}.

\subsection{Simulation Study}
\label{s:simulation}
Since the inverse temperature cannot be directly observed, we have used a simulation study to evaluate the accuracy of our method where the true value of $\beta$ is known. Following a similar methodology to that introduced by \citet{Cook2006}, we first simulated 20 values of $\beta$ from the prior, then generated 20 images from the model that corresponded to those parameter values. Each image had $125 \times 125$ pixels with $k=3$ unique labels. We used a uniform prior on the interval $[0, 1.005]$ for $\beta$ and natural conjugate priors $\pi(\mu_j) \sim \mathcal{N}(0, 100^2)$ and $\pi(\sigma_j^2) \sim \mathcal{IG}(1, 0.01)$ for the additive Gaussian noise.

Precomputation of the binding function took 1 hour 23 minutes for 987 values of $\beta$, using 1000 iterations of Swendsen-Wang for each. The resulting estimates of $\hat\mu(\beta)$ and $\hat\sigma(\beta)$ are illustrated in Fig.~\ref{f:simSz}. Total CPU time for all 16 parallel threads was 21 hours 50 minutes, indicating over 98\% utilisation of the available capacity. Memory usage was less than 1.3GB. Since the same mapping function was reused for all 20 images, this runtime could be amortised across the entire corpus. Thus, the cost of precomputation was less than 5 minutes per image.

\begin{figure}
\begin{center}
\includegraphics[width=\linewidth]{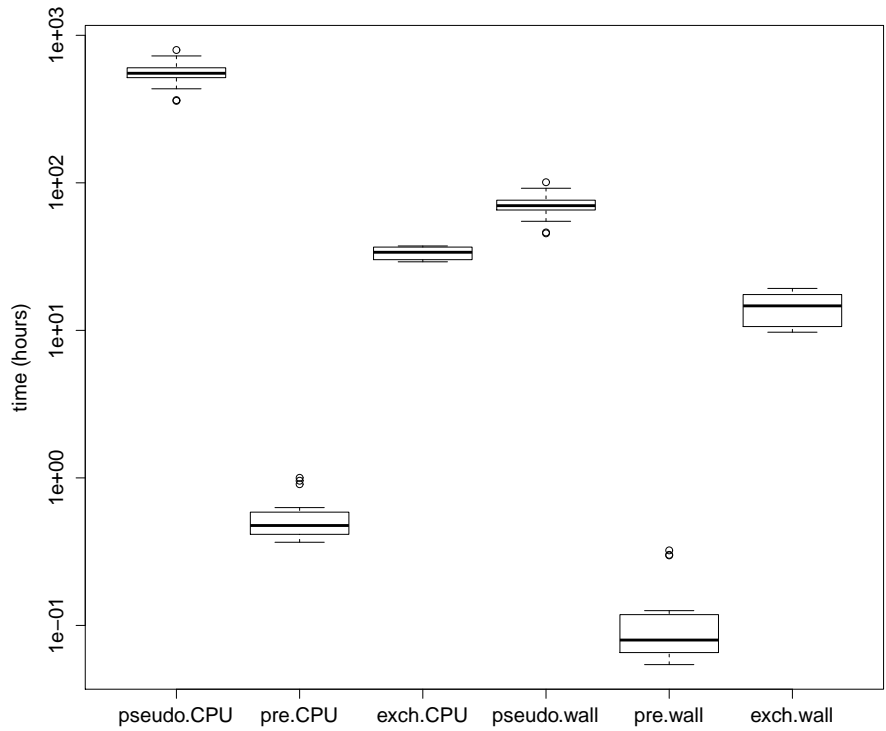}
\end{center}
\caption{Distribution of CPU times (left) and elapsed (wall clock) times for model fitting with the exchange algorithm in comparison to SMC-ABC with pseudo-data or precomputed $\hat\phi(\beta)$.}
\label{f:runtime}
\end{figure}
Across the 20 simulated images, our algorithm took an average of 7 minutes per image for between 49 and 107 SMC iterations. Using pseudo-data to compute the sufficient statistic took an average of 71.4 hours for 45 to 111 SMC iterations. The approximate exchange algorithm took an average of 14.4 hours for 100,000 iterations. Fig.~\ref{f:runtime} illustrates that this two orders of magnitude difference in the distribution is consistent for both elapsed (wall clock) time and CPU time. This shows that the gain in performance is due to computational efficiency, not because of any increase in parallelism.

\begin{figure*}
  \subfloat[SMC-ABC with pseudo-data]{\label{f:sim_post_ABC-SMC} \includegraphics[scale=0.25]{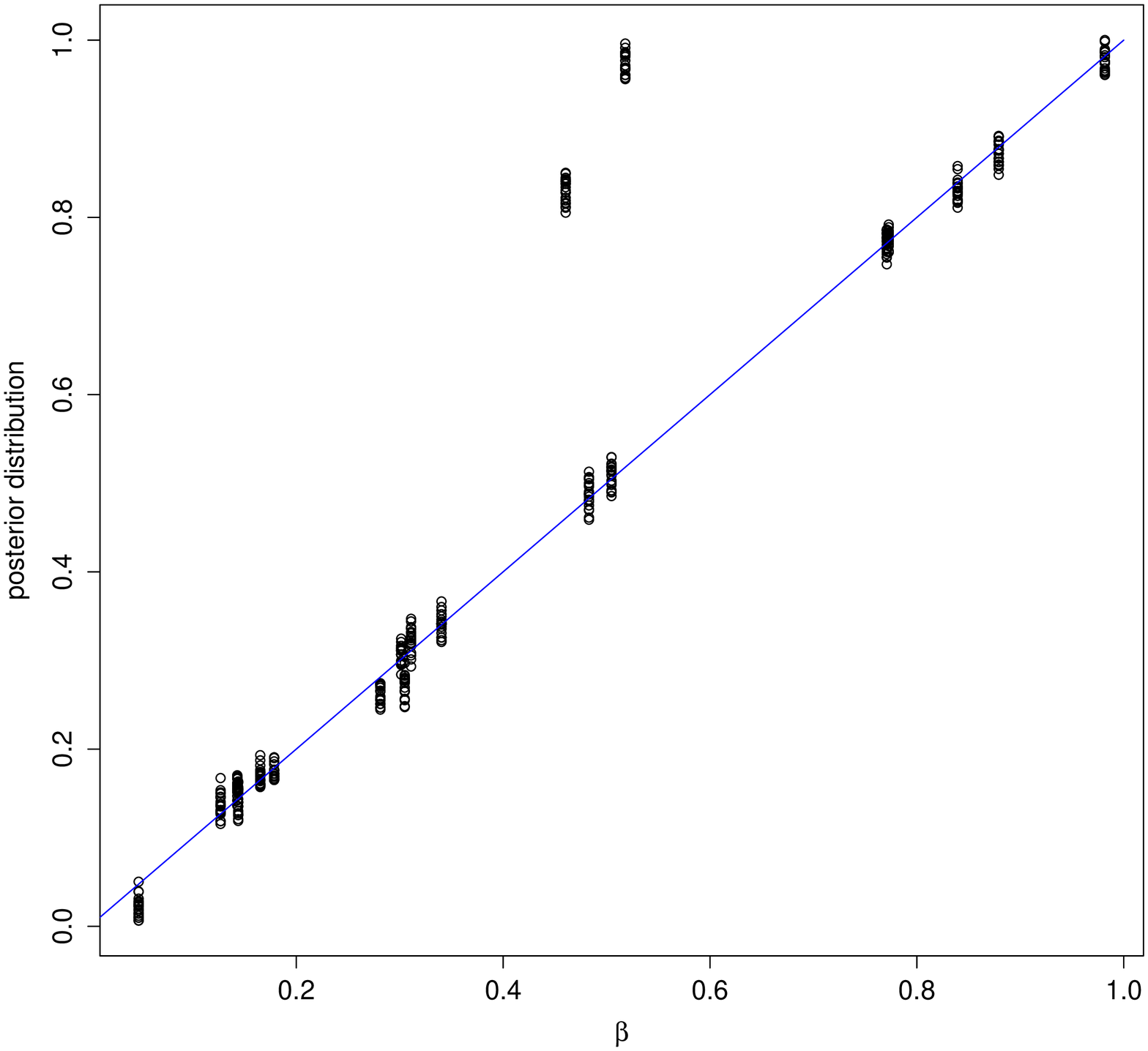}}
\qquad
  \subfloat[SMC-ABC with pre-computation]{\label{f:sim_post_ABCpre} \includegraphics[scale=0.25]{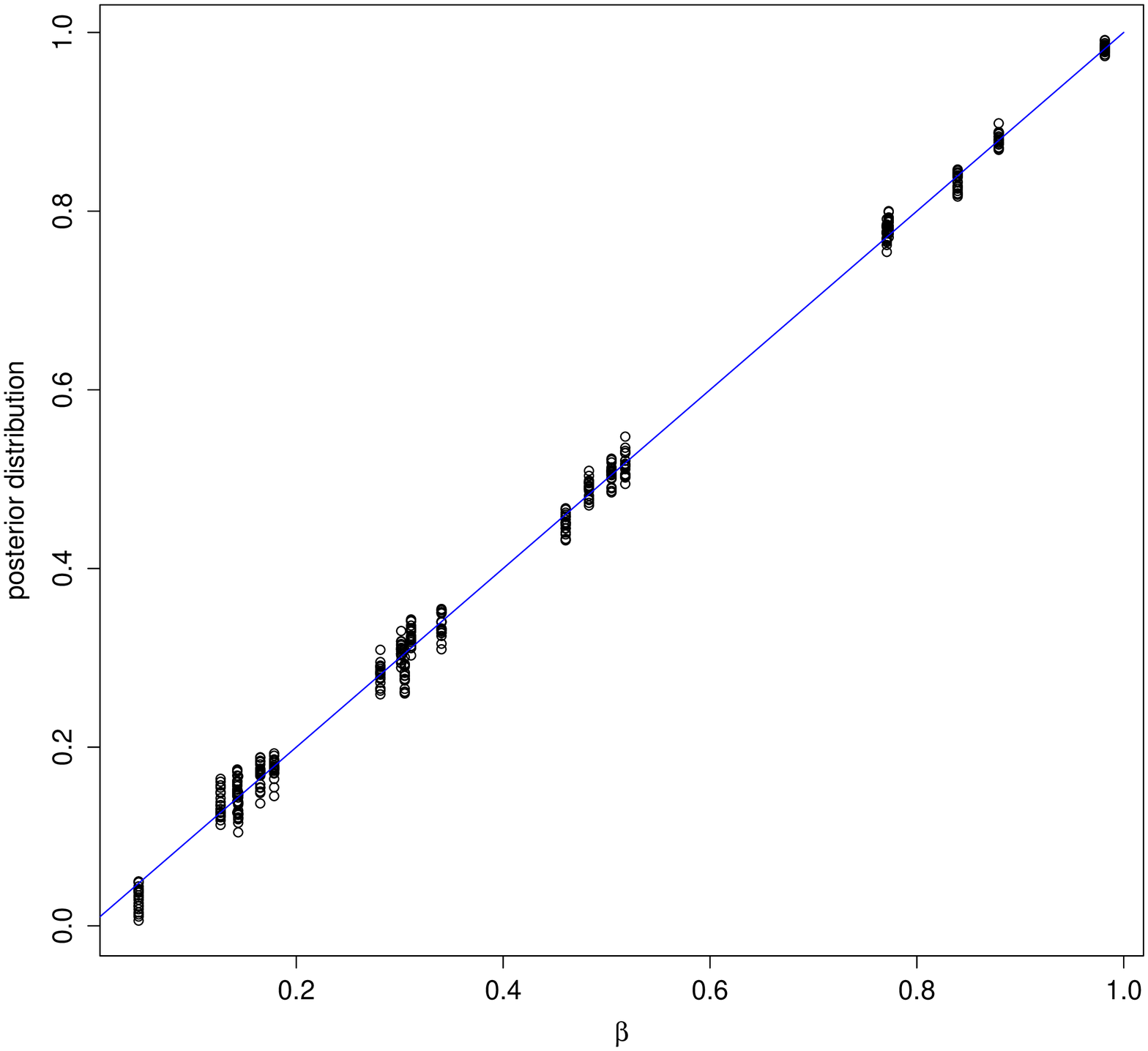}}
\qquad
  \subfloat[approximate exchange algorithm]{\label{f:sim_post_exch} \includegraphics[scale=0.25]{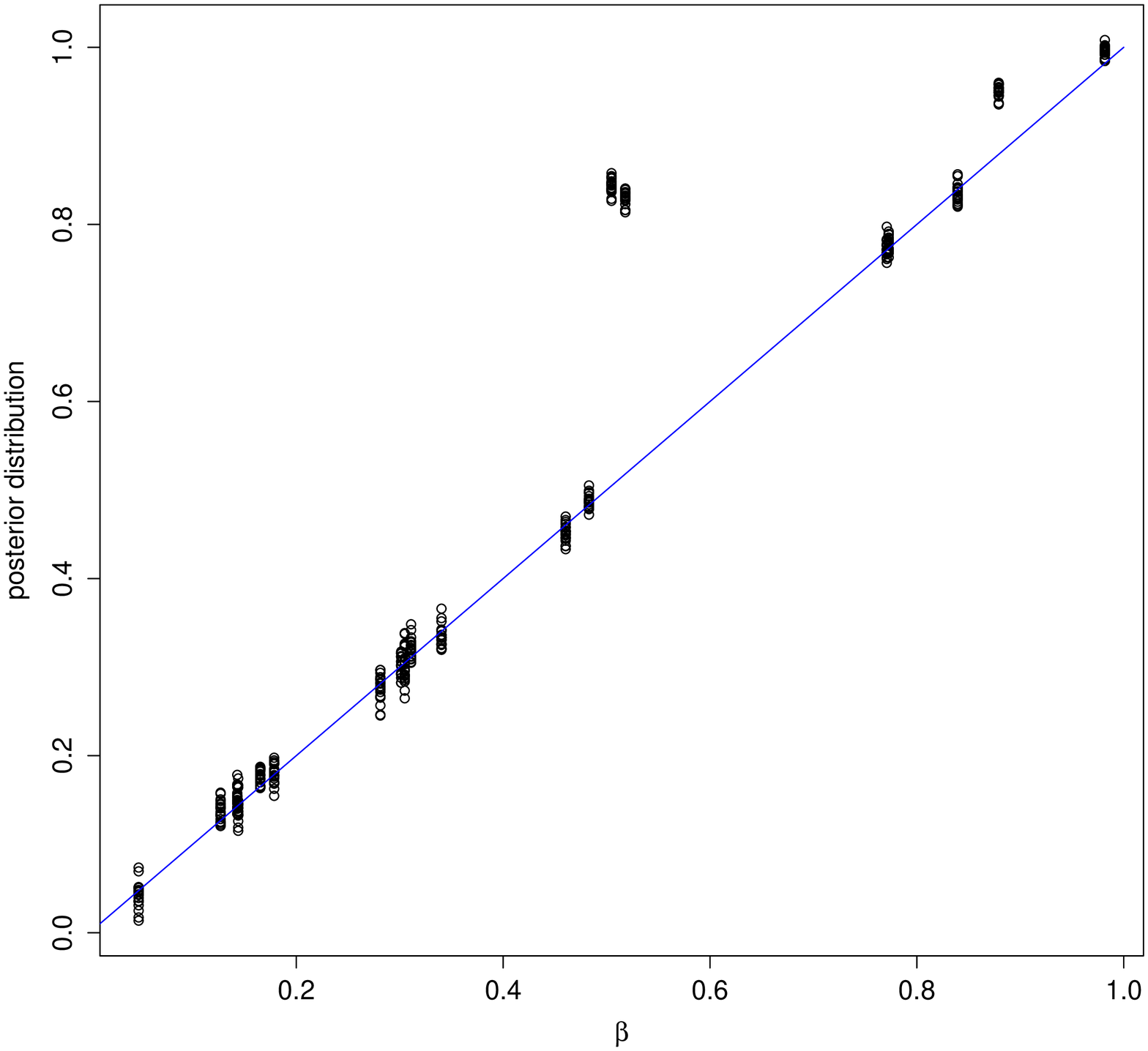}}
\caption{Results for the simulation study of 20 images. The $x$ axis is the true value of $\beta$ and the $y$ axis shows the posterior samples from $\pi_t(\beta|\mathbf{z})$}
\label{f:sim_results}
\end{figure*}
Fig.~\ref{f:sim_results} shows that both SMC-ABC and the approximate exchange algorithm produced erroneous estimates of $\beta$ for some of the images, with a large difference between the posterior distribution of the particles and the true value of the inverse temperature. This is most likely due to problems with the Markov chain undergoing phase transition, as explained in section~\ref{s:noise}. We were able to compensate for this problem by increasing the number of replicates of the summary statistic to $M=200$. Thanks to precomputation of the binding function, this had little impact on the runtime of our algorithm.

\subsection{Satellite Remote Sensing}
\label{s:satellite}
We have also illustrated our method on real data, using a satellite image of Brisbane, Australia. The pixel values correspond to the Normalized Difference Vegetation Index (NDVI), which was calculated accordng to:
\begin{equation}
\label{eq:ndvi}
\mathrm{NDVI} = \frac{NIR - VIS}{NIR + VIS}
 \end{equation}
In Landsat 7 images, the visible red ($VIS$) band corresponds to light wavelengths of $(0.63 \dots 0.69){\mu}m$ and the near-infrared ($NIR$) band corresponds to $(0.76 \dots 0.90){\mu}m$~\citep{NASA2011}.

The image was cropped to a region of interest that was approximately 40km east to west and 20km north to south, containing a total of 978,380 pixels. This region includes the city centre as well as suburbs and national parks to the south and west. By labelling the pixels, we aim to quantify the levels of vegetation present in the area and identify contiguous clusters of forest and parkland.

 Precomputation of the mapping function took 13 hours 23 minutes for 987 values of $\beta$. Total CPU time for all 16 parallel threads was 75 hours 40 minutes, indicating over 98\% utilisation of the available capacity. Memory usage was approximately 6.5GB.

We used weakly informative priors $\pi(\beta) \sim \mathcal{U}[0, 3]$, $\pi(\mu_j) \sim \mathcal{N}(\bar{\vec{y}}, 5)$, and $\pi(\sigma_j^2) \sim \mathcal{IG}(1, 0.01)$ for the hidden Potts model with $k=6$ unique labels. Model fitting took 5 hours 36 minutes using our algorithm. CPU time for 8 parallel threads was 39 hours, indicating 88\% utilisation. The 95\% posterior credible interval for $\beta$ was $[1.243; 1.282]$.

Running the original SMC-ABC algorithm of \citet{DelMoral2012} on this dataset is clearly infeasible, due to the cost of simulating pseudo-data. It takes 89 hours to perform a single SMC iteration on our hardware. We also found that the exchange algorithm of \citet{Murray2006} was unable to scale to data of this dimension, even when using 500 iterations of Gibbs sampling for the auxiliary variable as recommended by~\citet{Cucala2009}. It took 97 hours for 10,000 MCMC iterations. Discarding the first 5,000 as burn-in left an effective sample size of only 390 due to auto-correlation of the Markov chain.

\section{Discussion}
\label{s:conclusion}
We have demonstrated that the scalability of ABC (and SMC-ABC in particular) can be dramatically improved using a precomputed binding function for indirect inference.  We observed two orders of magnitude improvement in runtime in our simulation study, from an average of 71 hours down to only 7 minutes. This enables Bayesian inference for datasets of realistic scale, for which it was previously infeasible. An important example is imaging datasets with a million pixels or more, such as satellite remote sensing. We showed that our algorithm was able to estimate the smoothing parameter of a satellite image, while neither the approximate exchange algorithm nor the SMC-ABC algorithm of \citet{DelMoral2012} were scalable enough to be practical.

The computation of the binding function is embarrasingly parallel and therefore can make full use of modern computer architectures. Once the binding function has been computed, it can be reused to fit the same model to multiple datasets. This is an advantage in many applications such as satellite imaging, where there are a large number of images with approximately the same dimensions. In a longitudinal setting it would also be possible to update the binding function sequentially as each image is processed, to increase its resolution in the region of highest posterior density.

The other major issue that we have addressed is additive Gaussian noise, which is commonly encountered in real world imaging data. \citet{Everitt2012} and \citet{Stoehr2014} have previously looked at ABC for latent models, where the state vector is not directly observed. However, their methods are only applicable where the observed data and the latent model are both discrete, sharing the same state space. The lack of identifiability induced by continuous observations creates a major problem for ABC, since the summary statistics become a moving target. We have introduced a pragmatic approach to reduce the tendency to become stuck in low-probability regions of the parameter space, while preserving the scalability of our method.

Our use of a precomputed binding function appears to be similar to that recently and independently developed in \citet{Cabras2014}, who select a multivariate normal auxiliary model for the summary statistic and also use a precomputation step similar to above.  It is important to note, however, that \citet{Cabras2014} assume that the covariance matrix of the auxiliary model is independent of $\theta$.  Such an assumption is severely violated in our Potts model application (see Fig.~\ref{f:exact_var_n}) and unlikely to hold generally across models with intractable likelihoods.  Furthermore, \citet{Cabras2014} assume throughout their paper that the summary statistic must be the same dimension as the parameter.  We note here that this assumption is not required to use the preprocessing idea detailed in this paper.  Finally, \citet{Cabras2014} use a regular grid over the parameter space. We suggest that for nonlinear, heteroskedastic auxiliary models it may be more appropriate to select a non-regular grid in order to obtain a good estimate of the mapping by including more points around regions where the gradient of the relationship is large.  

The degree of complexity required for the binding function is dependent on the dimensionality of the parameter space, the number of summary statistics, and the properties of the relationship between them. These factors will also influence how much pseudo-data (in terms of the grid size and the number of replicate datasets) must be simulated in order to achieve a sufficiently good approximation of the likelihood. The nonlinear, heteroskedastic regression that was applied by \citet{Blum2010a} or \citet{Wilkinson2014} would be a good choice in many cases, although simpler techniques for estimating the binding function could also be used. We assume that the summary statistics can be modelled as a continuous and smoothly-varying function of the parameters. The output of the precomputation step can be used to verify empirically that this assumption holds for the specific model under consideration.

\citet{Cabras2014} have shown that ABC with indirect inference can also be applied to multivariate likelihoods, but only where the assumption of homoscedasticity is met. More research is needed to extend this method to Gibbs random fields such as the exponential random graph model (ERGM), which exhibits correlation between its summary statistics dependent on the value of $\theta$. 

The current methods for indirect inference would suffer from the curse of dimensionality if applied to models with a huge number of parameters and/or summary statistics, such as those encountered in genetics. Our method degrades very quickly as the number of parameters grows, because of a multi-grid requirement that grows as a power of the dimension of the parameter. We rely on asymptotic arguments as  $M \rightarrow \infty$, therefore obtaining a sufficient number of simulations will become much more difficult as the dimension of the parameter space increases.
 The method that we have described is not universally applicable, nevertheless there exist a wide variety of models to which it could be successfully applied.

\section*{Acknowledgements}
The authors would like to thank the organisers and attendees of the MCMSki conference for their interest and feedback. In particular, we are grateful to D. P. Simpson, A. Mira, and the anonymous reviewers for their thoughtful comments and suggestions on an earlier version of this manuscript. M. T. Moores acknowledges the financial support of Queensland University of Technology and the Australian federal government Department of Education, Science and Training. C.P. Robert's research is supported by the Agence Nationale de la Recherche (ANR 2011 BS01 010 01 projet Calibration) and an Institut Universitaire de France senior grant 2010-2016. K. L. Mengersen's research is funded by a Discovery Project grant from the Australian Research Council. Landsat imagery courtesy of NASA Goddard Space Flight Center and U.S. Geological Survey. Computational resources and services used in this work were provided by the HPC and Research Support Group, Queensland University of Technology, Brisbane, Australia.

\bibliographystyle{plainnat}
\bibliography{abc}

\end{document}